\input harvmac
\input epsf

%
%

\nref
\CremmerI{ E. Cremmer, B Julia and J. Scherk, ``Supergravity Theory
in 11 Dimensions," Physics Letters Volume 76B, Number 4
( 409 - 412 ) }

\nref
\DavisI{ K. Davis, ``M-Theory and String-String Duality,"
hep-th/9601102. }

\nref
\DasguptaI{K. Dasgupta and S. Mukhi, ``Orbifolds of M-Theory,"
hep-th/9512196; TIFR/TH/95-64; ( To appear in Nuclear Physics B ) }

\nref
\DuffII{ M.J. Duff, R.R. Khuri, and J.X. Lu, ``String Solitons,"
hep-th/9412184; Physics Reports {\bf 259} ( 1995 ) 213-326 }

\nref
\GinspargI{ P. Ginsparg, ``On Torodial Compactification of Heterotic
Superstrings," Physical Review D Volume 35, Number 2 ( 648 - 654 ) }

\nref
\HullI{C.M. Hull and P.K. Townsend,``Unity of Superstring
Dualities," hep-th/9410167; QMW-94-30; R/94/33; Nuclear Physics
B {\bf 438} (1995) 109 - 137 }

\nref
\SchwarzI{J.H. Schwarz, ``M-Theory Extensions of T-Duality,"
hep-th/9601077, CALT-68-2034. }

\nref
\SeibergI{M. Dine, P. Huet and N. Seiberg, ``Large and Small
Radius in String Theory," Nuclear Physics B {\bf 322} (1989) 301-316 }

\nref
\SenI{A. Sen, ``String String Duality Conjecture in Six Dimensions
and Charged Solitonic Strings," hep-th/9504027, TIFR-TH-95-16;
Nuclear Physics B {\bf 450} (1995) 103-114 }

\nref
\StrommingerI{ A. Stromminger, ``Massless Black Holes and Conifolds
in String Theory," hep-th/9504090; Nuclear Physics B {\bf 451} ( 1995 )
96-108 }

\nref
\VafaI{ C. Vafa, ``Evidence for F-Theory," hep-th/9602022,
HUTP-96/A004 }

\nref
\VafaII{ D. Morrison and C. Vafa, ``Compactification of F-Theory on
Calabi-Yau Threefolds I," hep-th/9602114; ( To appear in Nuclear
Physics B ) }

\nref
\VafaIII{ D.Morrison and C. Vafa, ``Compactification of F-Theory on
Calabi-Yau Threefolds II," hep-th/9603161 }

\nref
\WittenI{E. Witten, ``String Theory Dynamics in Various
Dimensions," hep-th/9503124; IASSNS-HEP-95-18;
Nuclear Physics B {\bf 443} (1995) 85 - 126  }

\nref
\WittenII{ E. Witten, ``Some Comments on String Dynamics,"
hep-th/9507121 }

\nref
\WittenIII{P. Horava and E. Witten, ``Heterotic and Type I String
Dynamics from Eleven Dimensions," hep-th/9510209;
IASSNS-HEP-95-86; PUPT-1571 }

\nref
\WittenVI{M.B. Green, J.H. Schwarz and E. Witten, (1987)
{\it Superstring Theory: Introduction } Cambridge :
Cambridge University Press }

\nref
\WittenVII{ E. Witten, ``Five-Branes and M-Theory on an Orbifold,"
hep-th/9512219; IASSNS-HEP-96-01}

\nref
\WittenVIII{ E. Witten and L. Alvarez-Gaume, ``Gravitational
Anomalies," Nuclear Physics B {\bf 234} (1983) 269-330 }

\nref
\WittenIX{ E. Witten , ``Global Gravitational Anomalies,"
Communications in Mathematical Physics {\bf 100} 197-229 (1985) }

%
%

\Title{
	\vbox{
		\baselineskip12pt
		\hbox{ RU 96-xx }
		\hbox{ hep-th/9608113 }
		\hbox{ Rutgers Theory }
		}
	}
	{
	\vbox{
		\centerline{ ENHANCED GAUGE SYMMETRY }
		\vskip 8pt
		\centerline{ IN }
		\vskip 8pt
		\centerline{ M-THEORY }
		}
	}

\centerline{ Kelly Jay Davis }
\bigskip
\centerline{Rutgers University}
\centerline{Serin Physics Laboratory}
\centerline{Piscataway, NJ 08855}

%
%

\vskip .3in
\centerline{ABSTRACT}
In this article we examine some points in the moduli  space of
M-Theory at which there arise enhanced gauge symmetries. In
particular, we examine the ``trivial" points of enhanced gauge
symmetry in the moduli space of M-Theory on $S^{ 1 } \times
S^{ 1 } / { \bf Z }_{ 2 }$ as well as the points of enhanced
gauge symmetry in the moduli space of M-Theory  on $ K3 $ and
those in the moduli space of M-Theory on $T^{5}/{\bf Z}_{2}
\times S^{1}$. Also, we employ the above enhanced gauge
symmetries to derive the existence of some points of enhanced
gauge symmetry in the moduli space of the Type IIA string
theory.
\Date{8/5/96}

%
%

\newsec{
		Introduction
	}

In the past year and a half much has changed in the field of string
theory. Early results relating the two Type II string theories
\SeibergI\ and the two Heterotic string theories \GinspargI\ have
been combined with newer results relating the Type II string theory
and the Heterotic string theory \SenI \WittenI\  as well as the
Type I string theory and the Heterotic string theory \WittenIII\ to
obtain a single ``String Theory." Also, there has been very much
progress in interpreting some, if not all, properties of String Theory
in terms of M-Theory \DavisI \DasguptaI  \HullI \SchwarzI \WittenI
\WittenIII \WittenVII. In addition recent progress
has been made in interpreting the properties of String Theory and
M-Theory in terms of F-Theory \VafaI \VafaII
\VafaIII. In this article, as I am ``morally" opposed to
breaking Lorentz symmetry, we will not consider F-Theory, but
we will examine some points of enhanced gauge symmetry which
arise in the moduli space of M-Theory compactified on various
manifolds. In addition, we will be examining the implications that
these points of enhanced gauge symmetry have for String Theory.
In doing so we will derive the existence of some standard
points of enhanced gauge symmetry in the Type IIA
moduli space. So, let us start by giving an overview of
the various relations which we will explore in this paper.

First and foremost, we should note a standard relation
\WittenVI\ which we will employ numerous times in this paper. If a
Heterotic string is compactified on a circle $S^{1}$, then it has a
relatively simple moduli space\foot{ Not, of course,
including internal radii.} which is simply $[ 0, \infty ]$,
corresponding to the radius of the $S^{1}$ upon which the Heterotic
string is compactified. At generic points in this moduli space, the
Heterotic theory possess a $U(1) \times U(1)$ gauge symmetry
corresponding to the two one-forms introduced by the compactification
of the ten-metric and ten-two-form. However, at a special radius $R =
\sqrt{ \alpha' }$, the so-called ``self-dual" radius, the $U(1) \times U(1)$
gauge symmetry is enhanced to an $SU(2) \times SU(2)$ gauge symmetry.
We will combine this standard relation with the more complicated
dualities between String Theory and M-Theory as well as dualities among
the various string theories to examine some points of enhanced gauge
symmetry which arise in the moduli space of M-Theory as well as
points of enhanced gauge symmetry which arise in the Type IIA moduli
space.

We will first examine some relatively standard points of enhanced
gauge symmetry in the moduli space of M-Theory. We will then move
onto less standard points of enhanced gauge symmetry. We will
first, as a ``warm-up," take a brief look at the points of
enhanced gauge symmetry in the moduli space
of M-Theory on $S^{1} \times S^{1} / {\bf Z}_{2}$. The
existence of these points of enhanced gauge symmetry follows
almost directly from the points of enhanced gauge symmetry
in the moduli space of the Heterotic
string on $S^{1}$. Witten and Horava  \WittenIII\  found that
M-Theory on $S^{1} / Z_{2}$ is equivalent to the $E_{8} \times E_{8}$
Heterotic string theory. Hence, as direct result, M-Theory on
$S^{1} \times S^{1} / {\bf Z}_{2}$ is equivalent to the Heterotic
string theory on $S^{1}$. Thus, M-Theory on $S^{1} \times S^{1} /
{\bf Z}_{2}$, as a result of the points of enhanced gauge symmetry
present in the moduli space of the Heterotic string theory on $S^{1}$,
possess an enhanced gauge symmetry at the self-dual radius of the
$S^{1}$ factor in  $S^{1} \times S^{1} / {\bf Z}_{2}$.

Next we will examine the points of enhanced gauge symmetry present
in the moduli space of M-Theory compactified on $K3$. Witten
\WittenI\ has shown that M-Theory on $K3$ is equivalent to the
Heterotic string theory on $T^{3}$. Now, as we previously mentioned,
the Heterotic string theory on $S^{1}$ possess an enhanced gauge
symmetry at a particular value of the $S^{1}$ radius. Similarly,
the Heterotic string theory on $T^{3}$ possess various points of
enhanced gauge symmetry corresponding to the various radii of the
three-torus $T^{3}$ becoming self-dual. Hence, M-Theory on $K3$
should also possess various points of enhanced gauge symmetry in
its moduli space corresponding to varying the parameters which
define the $K3$ compactification. In addition, we will examine what
the points of enhanced gauge symmetry in the moduli space of
M-Theory on $K3$ imply about enhanced gauge symmetries of the
Type IIA theory on $K3$.

Finally, we will examine the points of enhanced gauge symmetry
present in the moduli space of M-Theory compactified on
$T^{5} / {\bf Z}_{2} \times S^{1}$. As shown by Dasgupta, Mukhi, and
Witten \DasguptaI \WittenVII, M-Theory on $T^{5} / {\bf Z}_{2}
\times S^{1}$ is equivalent to the Type IIB string theory on
$K3 \times S^{1}$. Furthermore, Seiberg, Dine, and Huet \SeibergI\
long ago showed that  the Type IIB string theory on $K3 \times S^{1}$
is equivalent to the Type IIA string theory on $K3 \times S^{1}$,
where a T-Duality transformation is performed on the final
$S^{1}$ factor. Also, by way of six-dimensional string-string duality
\SenI \WittenI, the Type IIA string theory on $K3 \times S^{1}$ is
equivalent to the Heterotic string theory on $T^{5}$. Hence,
chaining together the above relations, M-Theory on
$T^{5} / {\bf Z}_{2} \times S^{1}$ is equivalent to the Heterotic
string theory on $T^{5}$. Thus, as varying the radii of the $T^{5}$
on the Heterotic side leads to enhanced gauge symmetries, the above
equivalence dictates that varying the parameters defining the
$T^{5} / {\bf Z}_{2} \times S^{1}$compactification leads to enhanced
gauge symmetries at points in the moduli space of M-Theory on
$T^{5} / {\bf Z}_{2} \times S^{1}$. We will then find that the
enhanced gauge symmetries at points in the moduli space of M-Theory
on $T^{5} / {\bf Z}_{2} \times S^{1}$ imply the existence of enhanced
gauge symmetries involving the Type IIA string theory on
$T^{5} / {\bf Z}_{2}$. With all this in mind let us start by looking
at the ``easy" case of M-Theory on $S^{1} \times S^{1} / {\bf Z}_{2}$.

%
%

\newsec{
		Enhanced Gauge Symmetries :
		M-Theory on $S^{1} \times S^{1} / {\bf Z}_{2}$
	}

In this section, as a ``warm-up" exercise, we will derive the
points of enhanced gauge symmetry present in the moduli space
of M-Theory on the manifold $S^{1} \times S^{1} / {\bf Z}_{2}$. To
do so, however, we will need to make use of the relation, derived
by Witten and Horava \WittenIII, between M-Theory on $S^{1} /
{\bf Z}_{2}$ and the $E_{8} \times E_{8}$ Heterotic string theory.
So, as a first step in our derivation, we will take a quick look at their
result.

\subsec{
		M-Theory on $S^{1} / {\bf Z}_{2}$
		$\sim$
		$E_{8} \times E_{8}$ Heterotic String Theory
	}

In this subsection we will quickly review the equivalence, established
by Witten and Horava \WittenIII, between M-Theory on $S^{1} /
{\bf Z}_{2}$ and the $E_{8} \times E_{8}$ Heterotic string theory.
So, let us start by considering the fact that the low-energy limit of
M-Theory is eleven-dimensional supergravity \WittenIII. Hence,
the low-energy limit of M-Theory employs a set of gamma matrices
$\Gamma^{I}$ which, as odd dimensions support only a
single chirality, satisfy the following relation
\eqn
\GammaMatrixEquationI{
\Gamma^{1} \Gamma^{2} \cdots \Gamma^{11} = 1.
}
\noindent Now, consider placing M-Theory on the manifold
$X_{10} \times S^{1}$, where $X_{10}$ is an arbitrary ten-manifold,
then modding the the theory by a ${\bf Z}_{2}$ which acts by a sign
change on the eleventh-coordinate $x_{11} \rightarrow -x_{11}$. In
this manner we obtain M-Theory on $X_{10} \times S^{1} / {\bf Z}_{2}$.
As the low-energy limit of M-Theory is eleven-dimensional supergravity,
M-Theory's low-energy limit is invariant with respect to a
supersymmetry generated by an arbitrary constant Majorana spinor
$\epsilon$. Dividing by the above ${\bf Z}_{2}$ rids us of half this
supersymmetry. If we choose our signs properly, we can require that
the supersymmetries generated by $\epsilon$ satisfy
\eqn
\GammaMatrixEquationII{
\Gamma^{11} \epsilon = \epsilon.
}
\noindent This, along with \GammaMatrixEquationI, implies that
$\epsilon$ satisfies
\eqn
\GammaMatrixEquationIII{
\Gamma^{1} \Gamma^{2} \cdots \Gamma^{10} \epsilon = \epsilon.
}
\noindent Hence, the supersymmetries not projected out by this
${\bf Z}_{2}$ action are chiral in a ten-dimensional sense.

Thus, the low-energy limit of M-Theory on $X_{10} \times S^{1} /
{\bf Z}_{2}$ reduces to a ten-dimensional theory with
one chiral supersymmetry. If M-Theory on $X_{10} \times S^{1} /
{\bf Z}_{2}$ is actually equivalent to a ten-dimensional
string theory, then this discovery limits our choices to the Type I
string theory or either of the two Heterotic string theories.
We will now show that if M-Theory on the manifold $X_{10} \times
S^{1} / {\bf Z}_{2}$ is equivalent to any string theory, then it is
the $E_{8} \times E_{8}$ Heterotic string theory on $X_{10}$.

We will do so by considering gravitational anomalies.
As shown long ago by Witten and Alvarez-Gaume \WittenVIII, the
only theories in which there may exist local gravitational anomalies
are those theories with spacetime dimension of the form  $4n + 2$,
for some integer $n$. Hence, there exist no local gravitational
anomalies in eleven-dimensions, as $11$ can not be written in the
form $4n + 2$ for some integer $n$. (One should also note that
Witten \WittenIX\ long ago established that global gravitational
anomalies may only exist in theories with spacetime dimension of the
form $8n + 1$ or $8n$; hence, there are no global gravitation
anomalies in eleven-dimensions.) So, with all these ``negative"
results in relation to eleven-dimensional gravitational anomalies,
we seem to be out of luck. But, this is actually not the case.
The fact that the manifold $X_{10} \times S^{1} / {\bf Z}_{2}$ has
a boundary actually comes to the rescue.

Consider the eleven-dimensional Rarita-Schwinger field and also think
of our compactification on $X_{10} \times S^{1} / {\bf Z}_{2}$ as a
compactification on $X_{10} \times S^{1}$ mod the ${\bf Z}_{2}$ action
$x_{11} \rightarrow -x_{11}$. Hence, the Rarita-Schwinger field may
be decomposed into its Fourier modes along the $S^{1}$ direction.
The non-zero Fourier modes along this direction have mass of order
$1 / R_{11}$, where $R_{11}$ is the radius of the $S^{1}$; hence, they
can not contribute to the anomaly. Only the zero mode may
contribute as it is generically massless. From a ten-dimensional
point-of-view, however, this massless mode leads to a chiral
ten-dimensional Rarita-Schwinger field. As $10$ is of the
form $4n + 2$ this ten-dimensional chiral Rarita-Schwinger field
can lead to an anomaly. The computation of this anomaly will
allow us to connect M-Theory and the $E_{8} \times E_{8}$
Heterotic string theory; so, let us proceed with its computation.

Consider an arbitrary diffeomorphisim of the eleven-dimensional
manifold $X_{10} \times S^{1} / {\bf Z}_{2}$ generated
by a vector $v^{I}$ such that $\delta x^{I} = \epsilon' v^{I}$.
The change in the effective action under this diffeomorphisim is
of the following form,
\eqn
\EffectiveActionVariationI{
	\delta \Gamma =
	i \epsilon' \int_{X_{10} \times S^{1} / {\bf Z}_{2}}
	d^{11}x \sqrt{g} v^{I} W_{I},
}
\noindent where $g$ is the eleven-dimensional metric and $W_{I}$ is
the anomalous variation in the effective Lagrangian, computable from
local data \WittenIII. Now, as we mentioned earlier, there are no
local gravitational anomalies in eleven-dimensions. Hence, the
contribution to the anomaly from a non-boundary point is zero and
the above integrand only has a non-zero support on the boundary of
$X_{10} \times S^{1} / {\bf Z}_{2}$. Let us notate the two boundary
components of the manifold $X_{10} \times S^{1} / {\bf Z}_{2}$ as
$A'$ and $A''$. Hence, the anomaly takes the form
\eqn
\EffectiveActionVariationII{
	\delta \Gamma =
	i \epsilon' \int_{ A' }
	d^{10}x \sqrt{g'} v^{I} W'_{I} +
	i \epsilon' \int_{ A'' }
	d^{10}x \sqrt{g''} v^{I} W''_{I},
}
\noindent where $g'$ and $g''$ are the restriction of $g$ to $A'$
and $A''$ respectively and $W'_{I}$ and $W''_{I}$ are constructed
from data at $A'$ and $A''$ respectively.

If we take the metric on $X_{10} \times S^{1} / {\bf Z}_{2}$
to be a standard metric on $S^{1} / {\bf Z}_{2}$ and an
arbitrary metric on $X_{10}$, then we can see that the above anomaly
is simply the standard gravitational anomaly in ten dimensions.
Hence, each boundary component of $X_{10} \times S^{1} / {\bf Z}_{2}$
contributes one-half of the standard ten-dimensional chiral
gravitational anomaly. Thus, as the anomaly is not zero, there must
be extra massless modes we do not yet know about living on the
boundary of $X_{10} \times S^{1} / {\bf Z}_{2}$. Also, as the
boundary is ten-dimensional, these massless boundary modes will
have to be vector multiplets as the ten dimensional vector multiplet
is the only ten-dimensional supermultiplet in which all particles
have spin less than 1 \WittenIII. So, one can practically
guess at this point, though a thorough justification \WittenIII\
requires a bit more work, that M-Theory on $X_{10} \times S^{1} /
{\bf Z}_{2}$ is related to the $E_{8} \times E_{8}$ Heterotic string
theory on $X_{10}$. As our above comments suggest, each
boundary component of $X_{10} \times S^{1}/{\bf Z}_{2}$
contributes one-half the standard ten-dimensional anomaly,
one would guess that one $E_{8}$ propagates $A'$
and the second $E_{8}$ propagates on $A''$. If we would have
considered either of the $N=1$ $SO(32)$ string theories, then we
would have been forced to place the entire $SO(32)$ multiplet on
$A'$ or $A''$ violating the above symmetry. Hence, as
sketched above, M-Theory on $X_{10} \times S^{1} /
{\bf Z}_{2}$ is equivalent to the $E_{8} \times E_{8}$
Heterotic string theory on $X_{10}$. Furthermore, aping arguments
put forth by Witten in \WittenI, we find that the $S^{1}/{\bf Z}_{2}$
radius $R_{11}$ as measured in the eleven-dimensional metric and the
ten-dimensional $E_{8} \times E_{8}$ Heterotic string theory
coupling constant $\lambda_{10,H}$ are related in the following 
manner
\eqn
\HeteroticMTheoryCouplingI{
	R_{11} = \lambda_{10,H}^{2 / 3}.
}
\noindent In addition, again aping the arguments put forth
by Witten in \WittenI, we find that the ten-dimensional
metric $G_{10,H}$ on $X_{10}$ given by the $E_{8} \times E_{8}$
Heterotic string theory is related to the ten-dimensional
metric $g_{10,M}$ on $X_{10}$ given by M-Theory in the
following manner
\eqn
\HeteroticMTheoryMetrics{
	g_{10,M} = R^{-1}_{11} G_{10,H}.
}
\noindent Now, let us employ the equivalence sketched above
along with \HeteroticMTheoryCouplingI\ and \HeteroticMTheoryMetrics\
to derive some information about the points of enhanced gauge
symmetry in the moduli space of M-Theory on
$S^{1} \times S^{1} / {\bf Z}_{2}$.

\subsec{
		M-Theory on $S^{1} \times S^{1} / {\bf Z}_{2}$ :
		Enhanced Gauge Symmetries
		}

In this subsection we will take a look at the points of enhanced
gauge symmetry in the moduli space of M-Theory on the manifold
$S^{1} \times S^{1} / {\bf Z}_{2}$. As a first step in this
direction, we should note that M-Theory on $S^{1} \times S^{1} /
{\bf Z}_{2}$, as sketched in the previous subsection, is equivalent
to the Heterotic string theory on $S^{1}$. Furthermore, as mentioned
in the introduction, the Heterotic string theory on a $S^{1}$ possess
a point of enhanced gauge symmetry in its moduli space. This point
occurs at the so-called self-dual radius of the $S^{1}$ on which
the Heterotic string theory is compactified. Taking $\alpha' = 1$,
this self-dual radius is given by $R^{2}_{10,H} = 1$, where
$R_{10,H}$ is the radius of the $S^{1}$ as measured in the
$E_{8} \times E_{8}$ Heterotic  string theory metric. Now, let us
interpret this in terms of M-Theory.

As mentioned in the previous subsection, the M-Theory metric $g_{10,M}$
on the ten-manifold $X_{10}$ and the $E_{8} \times E_{8}$ Heterotic
string theory metric $G_{10,H}$ on the same manifold $X_{10}$ are
related, $g_{10,M} = R^{-1}_{11} G_{10,H}$. As a result, distances measured
in the Heterotic string theory metric are scaled as compared to
distances measured in the M-Theory metric. In particular, any distance
$D_{H}$ in the Heterotic string theory metric is related to a the same
distance $D_{M}$ as measured in the M-Theory metric by $D_{H}
R^{-1/2}_{11} = D_{M}$. So, in particular, we may employ this relation to
interpret the self-dual radius of the Heterotic string theory in a
M-Theory context.

The self-dual radius in the Heterotic string theory occurs at
$R^{2}_{10,H} = 1$. If we multiply both sides of this equation by
$R^{-1}_{11}$ and employ the above scaling between M-Theory and
Heterotic theory distances, then we find that the points of enhanced
gauge symmetry in the M-Theory moduli space occur on the curve
\eqn
\MTheoryEnhancedPointsI{
	R^{2}_{10}R_{11} = 1,
	}
\noindent where $R_{10}$ is the $S^{1}$ radius as measured in the
M-Theory metric and $R_{11}$ is the $S^{1} / {\bf Z}_{2}$ radius as
measured in the M-Theory metric. A graphical representation of
this result is shown in Figure 1.

\centerline{
	\epsfbox{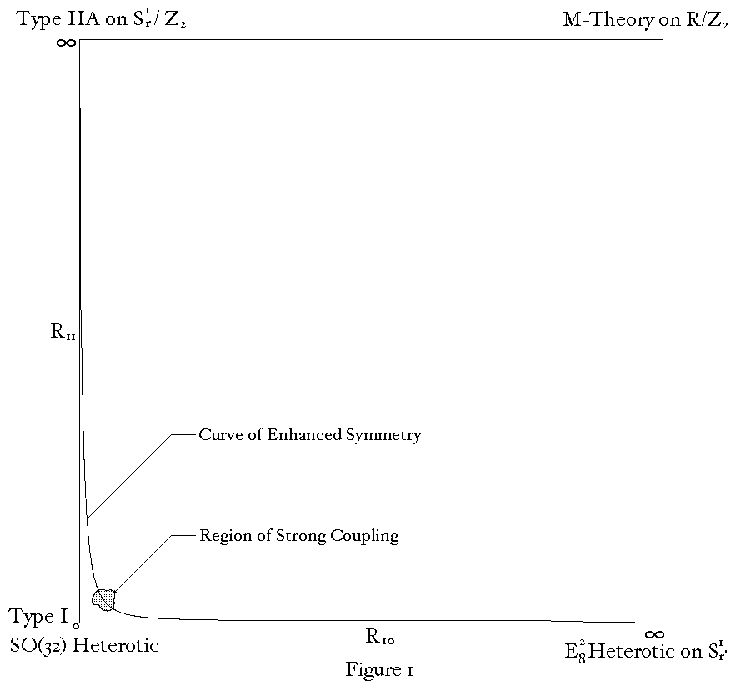}
}

As we have indicated in Figure 1, some regions in the moduli space
of M-Theory on $S^{1} \times S^{1} / {\bf Z}_{2}$ lend themselves
to a ``natural" interpretation in terms of various weakly coupled
string theories. Let us next take a moment to comment on the
various limits proposed in Figure 1.

Let us first consider the limit in which both $R_{10}$ and $R_{11}$
are taken to infinity in the M-Theory compactification on $S^{1}
\times S^{1} / Z_{2}$. Obviously, as $R_{10}$ corresponds to the
$S^{1}$ radius and $R_{11}$ to the $S^{1} / {\bf Z}_{2}$ radius,
the limit in which both $R_{10}$ and $R_{11}$ go to infinity
corresponds to M-Theory on ${\bf R} / {\bf Z}_{2}$. Now, let us
comment on the Type IIA string theory limit.

As Figure 1 indicates, the Type IIA string theory limit corresponds
to the limit in which $R_{11}$ goes to infinity and $R_{10}$ to zero.
Let us try and understand this a bit better. As was shown by Witten
\WittenI\ M-Theory compactified on an $S^{1}$ is equivalent to the
Type IIA string theory. This equivalence is similar to the
Heterotic/M-Theory equivalence we sketched above and the details of
its derivation are presented elsewhere \DavisI; so, we will not
review them here. As we are considering M-Theory on $S^{1} \times
S^{1} / {\bf Z}_{2}$, Witten's equivalence between M-Theory on
$S^{1}$ and the Type IIA string theory \WittenI\ implies that
M-Theory on $S^{1} \times S^{1} / {\bf Z}_{2}$ is equivalent to the
Type IIA string theory on $S^{1} / {\bf Z}_{2}$. Now, as shown by
Witten \WittenI, the ten-dimensional coupling constant of the Type
IIA string theory $\lambda_{10,IIA}$ and the radius $R_{10}$ of the
$S^{1}$ as measured in the M-Theory metric are related as follows
$R_{10} = \lambda^{2/3}_{10,IIA}$. Hence, the weak coupling limit
of the Type IIA string theory occurs when we take $R_{10}$ to zero,
as indicated in Figure 1. Also, as indicated in Figure 1, the limit
in which $R_{11}$ goes to infinity and $R_{10}$ to zero corresponds
to the Type IIA string theory on $S^{1}_{r} / {\bf Z}_{2}$. The
$S^{1}_{r} / {\bf Z}_{2}$ factor follows from the Type IIA string
theory interpretation of the limit $R_{11} \rightarrow \infty$.
As $R_{11}$ is measured in the M-Theory metric, we can not
automatically assume that taking the limit $R_{11} \rightarrow
\infty$, and thus $S^{1} / {\bf Z}_{2} \rightarrow {\bf R} /
{\bf Z}_{2}$ in the M-Theory picture, corresponds to a similar story
in the Type IIA string theory picture. We must translate this limit
to a limit taken in the Type IIA string theory metric.

This translation between the two limits, and hence the two metrics,
is analogous to the metric translation between the ten-dimensional
Heterotic string theory metric and the ten-dimensional M-Theory
metric put forth at the end of last subsection. In fact the
ten-dimensional Type IIA metric $G_{10,IIA}$ and the ten-dimensional
M-Theory metric $g_{10,M}$ are related by $g_{10,M} = R^{-1}_{10}
G_{10,IIA}$ \DavisI \WittenI. Hence, we find \DavisI \WittenI\
that the $S^{1} /{\bf Z}_{2}$ radius $R_{11}$, as measured in the
M-Theory metric, becomes $R^{1/2}_{10}R_{11}$ when measured in the
Type IIA metric. Hence, in the limit $R_{11} \rightarrow \infty$ and
$R_{10} \rightarrow 0$ the radius of the $S^{1} / {\bf Z}_{2}$
factor as measured in the Type IIA metric can take on any value
depending upon how one goes to infinity. In particular, consider
taking a variable $q$ to infinity $q \rightarrow \infty$, then
defining the limits of $R_{10}$ and $R_{11}$ by their relation to
$q$ such that $R_{11} = q$ and $R_{10} = r^{2}q^{-2}$. Doing so and
taking the limit as $q \rightarrow \infty$ one finds that the
$S^{1} / {\bf Z}_{2}$  radius as measured in the Type IIA metric is
simply $R^{1/2}_{10} R_{11} = r$. So, ``generically" we have found
that in the limit $R_{10} \rightarrow 0$ and $R_{11} \rightarrow
\infty$ the radius of $S^{1} / {\bf Z}_{2}$ as measured in the Type
IIA string theory metric is finite. However, we have also shown that
the ten-dimensional Type IIA string theory coupling constant
$\lambda_{10,IIA}$ goes to zero in the same limit. Hence,  at a
``generic," finite compactification radius of $S^{1} / {\bf Z}_{2}$,
the Type IIA string theory is a ``good" description of the limit
$R_{11} \rightarrow \infty$ and $R_{10} \rightarrow 0$ as it is
weakly coupled at its compactification radius, justifying its
inclusion in Figure 1. Next let us consider the limit in which
$R_{11} \rightarrow 0$ and $R_{10} \rightarrow \infty$.

As we sketched in the previous subsection, M-Theory on the manifold
$X_{10} \times S^{1} / {\bf Z}_{2}$ is equivalent to an $E_{8} \times
E_{8}$ Heterotic string theory on $X_{10}$. So, in particular,
M-Theory on $S^{1} \times S^{1} / {\bf Z}_{2}$ is equivalent to the
Heterotic string theory on $S^{1}$, where, in accord with
\HeteroticMTheoryCouplingI, the ten-dimensional coupling constant
$\lambda_{10,H}$ of the Heterotic string theory is related to
$R_{11}$ by $\lambda_{10,H} = R^{3/2}_{11}$. So, as we are
considering the limit in which $R_{11} \rightarrow 0$,
this limit corresponds to the weakly coupled Heterotic theory on
$S^{1}$, i.e. $\lambda_{10,H} \rightarrow 0$. Furthermore, as we are
also considering the limit in which $R_{10} \rightarrow \infty$, as
measured by in M-Theory metric, the $S^{1}$ factor of $S^{1} \times
S^{1} / {\bf Z}_{2}$ in this M-Theory limit becomes ${\bf R}$.
However, this limit is taken in the M-Theory metric and we must
interpret this limit in the metric of the Heterotic string theory.

We can do this by simply employing the relation between the
ten-dimensional metric of M-Theory and the ten-dimensional
metric of the Heterotic string theory presented in equation
\HeteroticMTheoryMetrics. This relation implies that the
$S^{1}$ radius $R_{10,H}$ as measured in the Heterotic theory is
$R_{10,H} = R^{1/2}_{11} R_{10}$. So, in the limit $R_{10}
\rightarrow \infty$ and $R_{11} \rightarrow 0$ the radius of the
$S^{1}$ factor in the M-Theory compactification on $S^{1} \times
S^{1} / {\bf Z}_{2}$ as measured in the Heterotic string theory
metric does not actually go to infinity, but can be taken to any
value depending upon the manner in which the limit is taken. For
instance, consider taking a variable $q$ to infinity $q \rightarrow
\infty$ and then defining $R_{11}$ by $R_{11} = r'{}^{2} q^{-2}$
and $R_{10}$ by $R_{10} = q$. In this case the $S^{1}$ radius as
measured in the Heterotic string theory metric is $R_{10,H} =
R^{1/2}_{11} R_{10} = r'$. So, as $r'$ is arbitrary the $S^{1}$
radius as measured in the Heterotic string theory metric can take
on any value in this limit. However, as we previously proved, the
ten-dimensional Heterotic string theory coupling constant
$\lambda_{10,H}$ in this limit goes to zero. So, at ``generic"
finite values of $r'$ the Heterotic string theory is weakly coupled
at its compactification radius and thus a good description of the
limit in which $R_{11} \rightarrow 0$ and $R_{10} \rightarrow \infty$.

Finally, let us look at the limit in which $R_{10} \rightarrow 0$
and $R_{11} \rightarrow 0$. In this limit, as the diagram suggests,
the theory can be interpreted in terms of a Type I string theory or
an $SO(32)$ Heterotic string theory. Let us see why this is the case.
As shown by Witten and Horava \WittenIII, the Type I string theory
and the $SO(32)$ Heterotic string theory are ``dual." Furthermore, as
established by Narin and clarified by Ginsparg \GinspargI, the $SO(32)$
and $E_{8} \times E_{8}$ Heterotic string theories are equivalent when
compactified on an $S^{1}$. So, employing these two relations
one may write the coupling constant and $S^{1}$ radius in
the limit $R_{10} \rightarrow 0$ and $R_{11} \rightarrow 0$ in
terms of the variables of an $SO(32)$ Heterotic string theory and/or a
Type I string theory. As the details of this derivation were presented
elsewhere \WittenIII, we will not repeat them here, but we will
simply quote the results\foot{ Note, the $R_{10}$ we are employing
is the $R_{11}$ of \WittenIII, and the $R_{11}$ of \WittenIII\
is our $R_{10}$ }. The ten-dimensional Type I coupling constant
$\lambda_{10,I}$ and the radius $R_{10,I}$ of the $S^{1}$ as
measured in the Type I metric are
\eqn
\TypeIRelationsI{
\eqalign{
&\lambda_{10,I} = { { R_{11} } \over { R_{10} } } , \cr
&R_{10,I} = { { 1 } \over { R_{11} R^{1/2}_{10} } }. \cr
}
}
\noindent Furthermore, we may also express the $SO(32)$
Heterotic coupling constant $\lambda_{10,H'}$ and the radius
$R_{10,H'}$ of the $S^{1}$ as measured in the $SO(32)$ Heterotic
metric in a similar manner,
\eqn
\HeteroticTwoCouplingConstantI{
\eqalign{
&	\lambda_{10,H'} = { { R_{10} }  \over { R_{11} } }, \cr
&	R_{10,H'} = { { 1 } \over { R_{10} R^{1/2}_{11} } }. \cr
}
}
\noindent Hence, in the limit $R_{10} \rightarrow 0$ and
$R_{11} \rightarrow 0$ one can see that the couplings
$\lambda_{10,I}$ and $\lambda_{10,H'}$ can be made
to go to finite values. Also, in this same limit one can
see that the radii $R_{10,I}$ and $R_{10,H'}$ both go to
infinity. Thus, ``generically," the $SO(32)$ Heterotic and
Type I string theories will be weakly coupled at their
compactification radii when $R_{10} \rightarrow 0$ and
$R_{11} \rightarrow 0$. However, one should note
that $\lambda_{10,I}$ and $\lambda_{10,H'}$ are
inverses of one another; thus, at any given point ``near"
the origin in Figure 1 one theory will a ``better"
description than the other. However, along the line
$R_{10} = R_{11}$ both the theories have coupling
order one and are equally ``bad" descriptions for
$R_{10,I}$ and $R_{10,H'}$ order one.

This leads us to an interesting region in the moduli
space of M-Theory on $S^{1} \times S^{1}/{\bf Z}_{2}$.
Notated by a ``cloud" in Figure 1, it is the region surrounding
the point where the line $R_{11} = R_{10}$ meets the curve
$R^{2}_{10} R_{11} = 1$. This is an interesting region in that
we have predicted an enhanced gauge symmetry should occur in
this region along $R^{2}_{10}R_{11} = 1$, but it seems we have
no perturbative ``handle" on the physics occurring in this
region. According to our above comments, both the Type I and
$SO(32)$ Heterotic theory have coupling constants and
compactification radii of order $1$ in this region. Similarly,
tracing our results for the Type IIA and $E_{8} \times E_{8}$
Heterotic string theories we find that both of these theories have
coupling constants and compactification radii of order $1$ in this
region. Hence, nothing seems to be weakly coupled at its
compactification radius in this region, but yet, strangely enough, we
can predict the existence of an enhanced gauge symmetry there. It
would be very interesting to try and further understand this region of
enhanced gauge symmetry from a perturbative view-point. Perhaps
F-Theory on $T^{2} \times S^{1} / {\bf Z}_{2}$ may come to the
rescue?

\newsec{
	Enhanced Gauge Symmetries :
	M-Theory on $K3$
	}

In this section we will examine the enhanced gauge symmetries
present in the moduli space of M-Theory on the four-manifold $K3$.
However, to do so we will have need of a relation, derived by
Witten \WittenI, between M-Theory on $K3$ and the Heterotic string
theory on $T^{3}$. So, we will first take a quick look at this
relation then later employ it to examine the points of enhanced
gauge symmetry in the moduli space of M-Theory on $K3$.

\subsec{
	M-Theory on $K3$
	$\sim$
	Heterotic String Theory on $T^{3}$.
	}

In this subsection we will briefly review the equivalence, established
by Witten \WittenI, between M-Theory on $K3$ and the Heterotic
string theory on $T^{3}$. The easiest manner in which to derive this
relation, and the one which we shall use,  employs string-string
duality in six-dimensions.

In six-dimensions there is a string-string duality \SenI \WittenI\
which states that the Heterotic string on $T^{4}$ is equivalent to
the Type IIA string on the manifold $K3$. From a slightly different
view-point we can consider this as an equivalence between the Type
IIA string theory on $K3$ and the Heterotic string theory on $T^{3}
\times S^{1}$. Furthermore, we may view the Heterotic string on
$T^{3} \times S^{1}$ as a Heterotic theory on $T^{3}$ compactified
on an $S^{1}$. From this final point-of-view we can consider the
seven-dimensional Heterotic coupling constant $\lambda_{7,H}$ and
the $S^{1}$ radius $R_{7,H}$ of $T^{3} \times S^{1}$ as measured in
the metric of the Heterotic string theory on $T^{3}$. These two
quantities are related in a simple manner to the six-dimensional
coupling constant $\lambda_{6,H}$ of the Heterotic string theory on
$T^{3} \times S^{1}$,
\eqn
\SixDSevenDCouplingRelation{
	{ { 1 } \over { \lambda^{2}_{6,H} } } =
	{ { R_{7,H} } \over { \lambda^{2}_{7,H} } }.
	}
\noindent Now, by way of six-dimensional string-string duality,
we may further relate $\lambda_{6,H}$, and thus $\lambda_{7,H}$,
to the Type IIA string theory coupling constant in six-dimensions
$\lambda_{6,IIA}$. According to standard string-string duality
in six-dimensions \DavisI \WittenI, $\lambda_{6,H}$ and
$\lambda_{6,IIA}$ are inverses of one another. Hence, we
have
\eqn
\SixDSixDCouplingRelation{
	\lambda_{6,IIA} =
	{ { 1 } \over { \lambda_{6,H} } } =
	{ { R^{1/2}_{7,H} } \over { \lambda_{7,H} } }.
	}
\noindent In addition, standard string-string duality \DavisI
\WittenI\ allows us to relate the Type IIA metric in six-dimensions
$G_{6,IIA}$ to the Heterotic metric $G_{6,H}$ in six-dimensions.
They are related by,
\eqn
\SixDSixDMetricRelation{
	G_{6,H} =
	\lambda^{2}_{6,H} G_{6,IIA} =
	{ { \lambda^{2}_{7,H} } \over { R_{7,H} } } G_{6,IIA},
}
\noindent where the second equality follows from
\SixDSevenDCouplingRelation.

Now, let us consider taking the limit in which the
$S^{1}$ factor of the Heterotic compactification on
$T^{3} \times S^{1}$ goes to ${\bf R}$, i.e. the limit in
which $R_{7,H} \rightarrow \infty$. In this limit
we will end up with the Heterotic string theory on $T^{3}$
and not $T^{3} \times S^{1}$; hence, this limit should,
if our conjectured equivalence, M-Theory on $K3$ $\sim$
Heterotic theory on $T^{3}$, holds, be equivalent to
M-Theory on $K3$. Let us show this is the case.

In the limit $R_{7,H} \rightarrow \infty$ the $S^{1}$
of the Heterotic compactification on $T^{3} \times S^{1}$
is becoming ``large." Hence, the states of the Heterotic
string which wind about the $S^{1}$ are becoming very
massive with mass of order $R_{7,H}$. Also, the states
of the Heterotic string which have a non-zero momentum
in the $S^{1}$ direction are becoming very light with
mass of order $1 / R_{7,H}$. In addition, as the three-torus
$T^{3}$ is not varying at all, the states with momentum and/or
windings about $T^{3}$ are unaffected in this limit.
As the Type IIA string theory on $K3$ is equivalent to the
Heterotic string theory on $T^{3} \times S^{1}$ \SenI \WittenI,
the limit $R_{7,H} \rightarrow \infty$ also may be interpreted
as a particular limiting case of the Type IIA string theory on
$K3$. Let us consider what this particular limit is.

As the Heterotic string theory possess a T-Duality symmetry
which exchanges the various radii of the $T^{4} = T^{3}
\times S^{1}$ compactification, there is not a unique map
from radii of the Heterotic compactification to the
parameters defining the Type IIA compactification.
However, one can choose a map, and this map will
be equivalent to a class of other maps by way of a T-Duality
transformation. So, with this in mind, let us choose
a map from the radii of the Heterotic string theory to the
parameters defining the compactification of the Type
IIA string theory on $K3$.

As we will show below, we may make a map from
the moduli space of the Heterotic string theory on
$T^{3} \times S^{1}$  to the moduli space of the
Type IIA string theory on $K3$ which maps the $S^{1}$
radius to the volume of the $K3$ factor. In particular,
if we denote the $K3$ volume as measured in the
ten-dimensional Type IIA string theory metric as
$V_{10,IIA}$, then the $S^{1}$ radius $R_{7,H}$ is
related to $V_{10,IIA}$ in the following manner,
\eqn
\VolumeRadiusRelationI{
	V_{10,IIA} =
	R^{2}_{7,H}.
}
\noindent Now, let us prove this relation is true.

Consider the Type IIA string theory in ten-dimensions. The
low-energy effective action for the Type IIA string theory is
Type IIA supergravity in ten-dimensions \WittenVI. The bosonic
fields present in Type IIA supergravity in ten-dimensions include
a metric $G_{10,IIA}$, a dilaton $\phi$, a one-form $A$, and a
three-form $A_{3}$. The bosonic portion of the low-energy
effective action for these fields is schematically \WittenI,
\eqn
\IIALowEnergyEffectiveActionTenD{
	\int d^{10} x
	\sqrt{ G_{10,IIA} }
	\left(
		e^{ - \phi } R +
		| dA |^{2} +
		| dA_{3} |^{2} +
		\cdots
	\right).
}
\noindent However, when we compactify this theory on $K3$
we get a set of various fields. In particular $A$
yields a six-dimensional one-form $a$ and no
zero-forms, $H^{1}(K3) = 0$. Also, $A_{3}$ yields
a six-dimensional three-form $a_{3}$ as well as
$22$ six-dimensional one-forms $C^{I}$, $H^{2}(K3)$
is $22$ dimensional. These all lead to a six-dimensional
action with the general form,
\eqn
\IIALowEnergyEffectiveActionSixD{
	\int d^{6}x
	\sqrt{ G_{6,IIA} }
	\left(
		{ {1} \over {\lambda^2_{6,IIA}} } R +
		V_{10,IIA} | da |^{2} +
		V_{10,IIA} | da_{3} |^{2} +
		| dC^{I} |^{2} +
		\cdots
	\right).
}
\noindent Now, in six-dimensions a three-form ``vector"
potential is dual to a one-form vector potential. Hence,
in the above action we may exchange the three-form
$a_{3}$ for a one-form ${\tilde a}$ where $V_{10,IIA}
da_{3} = * d{\tilde a}$.
Doing so we find,
\eqn
\IIADualLowEnergyEffectiveActionSixD{
	\int d^{6}x
	\sqrt{ G_{6,IIA} }
	\left(
		{ {1} \over {\lambda^2_{6,IIA}} } R +
		V_{10,IIA} | da |^{2} +
		V^{-1}_{10,IIA}  | d{\tilde a} |^{2} +
		| dC^{I} |^{2} +
		\cdots
	\right).
}
Now, to get a physical ``feel" for what the above action ``means" one
should consider the fact that the canonical action for a one-form
$A$ in six-dimensions with effective charge $e$ is,
\eqn
\CanonicalOneFormAction{
	\int d^{6}x
	{ {1} \over {4e^{2}} }
	| dA |^{2}.
}
\noindent Hence, in the action \IIADualLowEnergyEffectiveActionSixD\
we have a one-form $a$ with effective charge order $V^{-1/2}_{10,IIA}$.
Also, we have a one-form ${\tilde a}$ with effective charge order
$V^{1/2}_{10,IIA}$ and $22$ one-forms $C^{I}$ with effective charge
order $1$.

Now, as was found in \DavisI \WittenI, the mass of RR charged
particles with charge $e$ and coupling $\lambda_{6,IIA}$ is of order
$e / \lambda_{6,IIA}$. Hence, if we let $\lambda_{6,IIA}$ be constant
and take $V_{10,IIA} \rightarrow \infty$, then we see that the RR
particles charged with respect to the one-form $a$ have a mass which
goes to zero in this limit and RR particles charged with respect to
${\tilde a}$ have a mass which goes to infinity in this limit. This
is analogous to behavior we found for the Heterotic string on
$T^{3} \times S^{1}$ in the limit $R_{7,H} \rightarrow \infty$, and the
matching of these two behaviors will be the basis for our map from
the moduli space of the Heterotic string theory to the moduli space
of the Type IIA string theory.

In particular, if we look a the mass of lightest RR particles
in the $V_{10,IIA} \rightarrow \infty$ limit in the Type IIA
string theory on $K3$, then we find they are charged with
respect to the one-form $a$, and hence, we see that their mass
is given by
\eqn
\TypeIIAMassLimit{
	M_{6,IIA} =
	{ {1} \over {V^{1/2}_{10,IIA} \lambda_{6,IIA}} }.
}
\noindent Similarly, if we consider the Heterotic string theory on
$T^{3} \times S^{1}$ in the limit $R_{7,H} \rightarrow \infty$,
then we can see that the lightest states are those which have
a non-zero momentum about $S^{1}$. Their mass is of
order,
\eqn
\HeteroticMassLimit{
	M_{6,H} =
	{ {1} \over {R_{7,H}} }
}
\noindent So, if we wish to identify these two classes of particles
with vanishing mass, then we should identify $M_{6,IIA}$
and $M_{6,H}$. However, before doing so we should note that
the six-dimensional Type IIA metric is scaled relative to the
six-dimensional Heterotic metric in accord with
\SixDSixDMetricRelation. Hence, we must also scale the
masses before equating them. Doing so, we find,
\eqn
\SixDMassRelation{
	M_{6,H} =
	\lambda^{-1}_{6,H} M_{6,IIA} =
	\lambda_{6,IIA} M_{6,IIA},
}
\noindent where we have employed \SixDSixDCouplingRelation\
to write the second equality. Hence, \TypeIIAMassLimit,
\HeteroticMassLimit, and \SixDMassRelation\ together imply
\eqn
\VolumeRadiusRelationII{
	V_{10,IIA} =
	R^{2}_{7,H},
}
\noindent as was promised in \VolumeRadiusRelationI.

Now, as we have the relation \VolumeRadiusRelationII\
between $R^{2}_{7,H}$ and $V_{10,IIA}$, we can relate
the Heterotic string theory limit $R_{7,H} \rightarrow \infty$
to a limit in the Type IIA string theory. The limit in which the
Heterotic string theory on $T^{3} \times S^{1}$ has the radius
$R_{7,H}$ of the $S^{1}$ factor go to infinity corresponds,
by way of \VolumeRadiusRelationII, to the volume of the
$K3$ factor in the Type IIA compactification
going to infinity. Now, let us relate this to M-Theory.

As proven by Witten \WittenI, M-Theory on $X_{10} \times
S^{1}$ is equivalent to the Type IIA string theory on $X_{10}$.
So, in particular, Type IIA string theory on $K3$ is equivalent
to M-Theory on $K3 \times S^{1}$, and thus by way of
string-string duality \DavisI\ M-Theory on $K3 \times S^{1}$ is
equivalent to the Heterotic string theory  on $T^{3} \times S^{1}$.
Witten also proved \WittenI\ the M-Theory metric $g_{10,M}$
on $X_{10}$ and the Type IIA metric $G_{10,IIA}$ on $X_{10}$
are related by,
\eqn
\MetricRelation{
	g_{10,M} = \lambda^{-2/3}_{10,IIA} G_{10,IIA}.
}
\noindent So, as the M-Theory metric is a scaled
version of the Type IIA metric, we can see that if the
$K3$ volume goes to infinity as measured in the Type
IIA metric, then the $K3$ volume as measured in the M-Theory
metric need not go to infinity. To understand this let
us look at the relation between the Type IIA metric
and the M-Theory metric in a bit more detail.

Consider the fact that the Type IIA coupling constant
in ten-dimensions is related to the Type IIA coupling
constant $\lambda_{6,IIA}$ of the Type IIA string theory
on $K3$ by the standard ``compactification relation"
obtained by integrating over $K3$,
\eqn
\TypeIIASixDTenDCouplingRelation{
	{ {V_{10,IIA}} \over {\lambda^{2}_{10,IIA}} } =
	{ {1} \over {\lambda^{2}_{6,IIA}} }.
}
\noindent In turn, as a result of \SixDSixDCouplingRelation\
and \VolumeRadiusRelationII\ we have,
\eqn
\Chicken{
	\lambda_{10,IIA} =
	{ {R^{3/2}_{7,H}} \over {\lambda_{7,H}} }.
}
\noindent Now, as we know that the Type IIA
metric and the M-Theory metric are related as
in \MetricRelation, we see that the volume of
$K3$ in the M-Theory metric $V_{10,M}$ is actually
finite in the limit $R_{7,H} \rightarrow \infty$. We
have,
\eqn
\MTheoryKThreeVolume{
	V_{10,M} =
	\lambda^{-4/3}_{10,IIA} V_{10,IIA} =
	\lambda^{4/3}_{7,H}R^{-2}_{7,H} V_{10,IIA} =
	\lambda^{4/3}_{7,H},
}
\noindent where the first equality follows from \MetricRelation,
the second from \Chicken,
and the final from \VolumeRadiusRelationII.

Furthermore, Witten \WittenI\ also found that in the equivalence
between M-Theory on $X_{10} \times S^{1}$ and the Type IIA string
theory, the $S^{1}$ radius $R_{11}$ as measured in the
M-Theory metric is related to the Type IIA coupling constant
$\lambda_{10,IIA}$ in the following manner,
\eqn
\Fish{
	R_{11} = \lambda^{2/3}_{10,IIA}.
}
\noindent Employing \Chicken\
we have,
\eqn
\Fowl{
	R_{11} = { {R_{7,H}} \over {\lambda^{2/3}_{7,H}} }.
}
\noindent Hence, in the limit $R_{7,H} \rightarrow \infty$ for
finite $\lambda_{7,H}$, the radius $R_{11}$ of the $S^{1}$ in the
M-Theory compactification on $K3 \times S^{1}$ goes to infinity.

In summary, we have found that, by way of string-string
duality and the Type IIA/M-Theory equivalence, M-Theory
on $K3 \times S^{1}$ is equivalent to the Heterotic string theory
on $T^{3} \times S^{1}$. Furthermore, in the limit in which
the $S^{1}$ radius of the Heterotic compactification goes
to infinity, the above equivalence dictates that the $S^{1}$
radius of the M-Theory compactification also goes to infinity.
Thus, this limit establishes the equivalence between the
Heterotic string theory on $T^{3}$ and M-Theory on $K3$, as
promised at the beginning of this subsection.

\subsec{
	M-Theory on $K3$ :
	Enhanced Gauge Symmetries
	}

In this subsection we will examine the points of enhanced gauge
symmetry which occur in the moduli space of M-Theory
on $K3$. As we established in the previous subsection, M-Theory
on $K3$ is equivalent to the Heterotic string theory on $T^{3}$.
Hence, as the Heterotic string theory on $T^{3}$ possess various
points of enhanced gauge symmetry in its moduli space, M-Theory
on $K3$ must also possess various points of
enhanced gauge symmetry in its moduli space. So, let us now
examine these points of enhanced gauge symmetry.

Consider the Heterotic string theory compactified on $T^{3}$.
Generically, the three-torus $T^{3}$ is specified by six parameters,
three radii and three ``angles." As the ``angles" are not central
to our concerns here, we will ignore them by assuming that the
three-torus $T^{3}$ upon which the Heterotic string theory is
compactified is a ``right" torus and thus specified by giving
three radii. Let us notate these three radii, as measured in the
ten-dimensional metric of an un-compactified Heterotic string
theory, by $R_{i,H}$, where $i \in \{8,9,10 \}$. Now, with this
limitation imposed upon the Heterotic string theory moduli
space, let us consider the construction of a map from the points
of enhanced gauge symmetry in the Heterotic string theory moduli
space to the corresponding points of enhanced gauge symmetry
in the M-Theory moduli space. We will start this process by
ascertaining which states of M-Theory on $K3$ may give rise
to the gauge particles with which we are concerned.

M-Theory \WittenVII\ is a theory with two-branes and five-branes.
However, as is well-known from standard quantum field theory,
gauge fields are zero-branes. Hence, to obtain an enhanced
gauge symmetry in M-Theory on $K3$ we must exhibit a
zero-brane gauge field which results from the two-branes and
five-branes of M-Theory. The manner in which to do this was
first found by Stromminger \StrommingerI. One can take a $p$-brane,
say, and wrap it around a $q$-cycle of the compact manifold in
question, which is $K3$ in our case. This produces a 
$( p - q )$-brane in the non-compactified portion of the theory.
So, for instance, in our case we are starting with a two-brane and
a five-brane in eleven-dimensions. We can wrap the two-brane
or five-brane about any cycle in the homology ring of $K3$.
However, as our goal is to obtain zero-branes in the resultant
seven-dimensional theory, we should look for two-cycles and/or
five-cycles. But, $K3$ is four-dimensional; hence, it possess
no five-cycles. Our only hope then, is to look for two-cycles.
In fact, $K3$ has $22$ two-cycles \DavisI. So, there are plenty
of two-cycles around which we can wrap the M-Theory two-brane
to obtain a zero-brane in seven-dimensions. Next, let us consider
how this gives us or does not give us the massless gauge particles
for which we are looking.

Consider M-Theory compactified on $K3$. As shown above,
the ``wrapped" two-branes give rise to zero-branes in the resultant
seven-dimensional theory. The zero-branes in the resultant
seven-dimensional theory will have mass of order the area of the
two-cycle about which they are wrapped. In particular \WittenII,
the mass of these zero-branes is the tension of the two-brane times
the area of the two-cycle. Let us notate this by,
\eqn
\MTheoryMassOfZeroBraneI{
	M_{i,M} =
	T_{2} A_{i,M}
}
\noindent where $A_{i,M}$ is the area of the $i$-th two-cycle as
measured in the M-Theory metric, $T_{2}$ is the tension of the
two-brane, and $M_{i,M}$ is the mass of the resultant
zero-brane as measured in the M-Theory metric. Hence, as the
area of a given two-cycle goes to zero we obtain a massless
zero-brane in the seven-dimensional world. It is such massless
zero-branes which we will identify with the ``enhanced" gauge fields
of the Heterotic theory on $T^{3}$. But, to do so we will need
a few more details about both M-Theory on $K3$ and the
Heterotic theory on $T^{3}$.

In the Heterotic string theory on $T^{3}$ for a given $S^{1}$ of
radius $R_{i,H}$, there are states which wind about this $S^{1}$ and
have mass order $R_{i,H}$. Similarly, there are also non-winding states
which have a non-zero momentum along this $S^{1}$ and have mass of
order $1 / R_{i,H}$, and there are states which do not wind about the
$S^{1}$ and have no momentum along $S^{1}$. However, by way of
our general discussion above, the non-winding states are one-branes
in seven dimensions, and the winding states are zero-branes in
seven-dimensions. So, if we were to try and identify one of these
three types of states with the wrapped two-brane states of M-Theory,
we would try and identify the winding states with the wrapped two-brane
states. Furthermore, this identification agrees with now standard results
of string theory \WittenVI. In the Heterotic string theory on $T^{3}$,
the states which are becoming massless at the points of enhanced gauge
symmetry are Heterotic strings which wind and have momentum along
the $S^{1}$ which is becoming ``critical." Hence, as we wish to
identify the wrapped two-branes of M-Theory with the gauge
particles of the Heterotic string theory's enhanced gauge symmetry,
we will identify the wrapped M-Theory two-branes with the
Heterotic one-branes which are wrapped and have momentum
along an $S^{1}$.

As we have established on a conceptual level that we are identifying
the wrapped two-branes of M-Theory with the wrapped string states
of the Heterotic theory, let us now employ this information to
derive a relation between the coordinates on the moduli space of
M-Theory and the coordinates on the moduli space of the Heterotic
string theory. To do so, we will need to look at the mass relation
for the Heterotic one-branes which wrap about and have a
non-zero momentum along a given $S^{1}$. In particular,
consider an $S^{1}$ with radius $R_{i,H}$, then the mass of such states
is given by \WittenVI,
\eqn
\HeteroticGaugeParticleMasses{
	M_{i,H} =
	{ {1} \over {R_{i,H}} } + R_{i,H} - 2.
}
\noindent Now, to identify the coordinates parameterizing the
moduli space of M-Theory on $K3$ and those parameterizing
the moduli space of the Heterotic string theory on $T^{3}$ we
must identify $M_{i,M}$ and $M_{i,H}$. However, we should
be careful in doing so as the masses are measured in different
seven-dimensional metrics. Hence, to write an equality
expressing the relation of $M_{i,M}$ to $M_{i,H}$ we must
find a relation between the M-Theory seven-metric $g_{7,M}$
and the Heterotic seven-metric $G_{7,H}$. However, as our
line of inquiry does not require an equality, we will simply
satisfy ourselves with a proportionality. We have,
\eqn
\SevenDMassRelation{
	T_{2} A_{i,M} \sim
	\left(
		{ {1} \over {R_{i,H}} } + R_{i,H} - 2
	\right).
}
\noindent This is the map for which we have been looking,
the map from the moduli space of M-Theory on $K3$ to
the moduli space of the Heterotic string theory on $T^{3}$.
Of course one should note, as was the case in our identification
of $V_{10,IIA}$ and $R^{2}_{6,H}$, this identification does
not take into account the action of the T-Duality group
on the moduli of the Heterotic theory, and hence would be
changed greatly if one acted on the moduli with such a
transformation.

Now, at this point the observant reader may feel a bit cheated.
The $K3$ upon which M-Theory is compactified contains $22$
two-cycles. Hence, there exist $22$ ways of obtaining an
enhanced gauge symmetry when a two-cycle collapses. This
in-turn leads to $22$ different $M_{i,M}$'s. However, on
the Heterotic side we only had $3$ $M_{i,H}$'s corresponding
to the three different radii of $T^{3}$. So, it seems that
the M-Theory side contains may more ways of obtaining an
enhanced gauge symmetry than the Heterotic string theory
side does. This is, however, is not really the case.

The Heterotic string theory, in addition to the three radii
explicitly in $T^{3}$, contains $16$ internal radii. So,
in total we now have $19$ radii, but $19 \ne 22$. However,
any one of the three radii of $T^{3}$ give two enhanced gauge
symmetries, i.e. as one of the radii $R_{i,H}$ of $T^{3}$
becomes self-dual, there exist two $U(1)$'s which become
two $SU(2)$'s. Hence, they are double counted. Thus, on
the Heterotic side we now have $16 + 3 + 3 = 22$ ways
of obtaining an enhanced gauge symmetry. This matches
exactly with the $22$ two-cycles of $K3$ that may
collapse and give an enhanced gauge symmetry on the
M-Theory side. So, we were not really {\it cheating} in our
derivation, just holding our cards a little close to our
chest.

Next, let us see if we can learn anything new about
String Theory from the existence of these points of
enhanced gauge symmetry in the moduli space of M-Theory.

\subsec{
	Surprises?
	}

In this subsection we will quickly examine some implications
that the points of enhanced gauge symmetry in the moduli
space of M-Theory on $K3$ have upon String Theory.
In particular, we will examine the points of enhanced
gauge symmetry present in moduli space of the
Type IIA string theory on $K3$ which can be derived
from the points of enhanced gauge symmetry present in
the moduli space of M-Theory on $K3$.

Consider M-Theory on $K3$, as we saw above, when various
two-cycles in $K3$ go to zero area there arise various
points of enhanced gauge symmetry in the moduli space
of M-Theory on $K3$. Furthermore, as Witten proved
\WittenI, M-Theory on $S^{1}$ is equivalent to the Type IIA
string theory. Hence, M-Theory on $K3 \times S^{1}$ is
equivalent to Type IIA string theory on $K3$. Furthermore,
as there exist points of enhanced gauge symmetry in
the moduli space of M-Theory on $K3$, there also exist
points of enhanced gauge symmetry in the moduli
space of M-Theory on $K3 \times S^{1}$. In particular,
some of them arise in the same manner we discussed
earlier, a two-brane wraps about a collapsing two-cycle.
So, due the equivalence of M-Theory on $K3 \times S^{1}$
and the Type IIA string theory on $K3$, one should expect
points of enhanced gauge symmetry to arise in the moduli
space of the Type IIA string theory on $K3$. In fact these
symmetries do arise, as has been known \WittenII. Type IIA
string theory on $K3$ is equivalent to the Heterotic string theory
on $T^{4}$ \SenI. Hence, as there exist points of enhanced
gauge symmetry in the moduli space of the Heterotic
string theory on $T^{4}$ there should also exist corresponding
points in the moduli space of the Type IIA string theory on
$K3$.

The means by which some of the points of enhanced gauge symmetry
arise was established by Witten \WittenII. Witten found that some
of the points of enhanced gauge symmetry in the moduli space
of the Type IIA string theory on $K3$ arise from two-branes
of the Type IIA string theory wrapping about collapsing two-cycles
of $K3$, the exact explanation we put forward in the context
of M-Theory. Furthermore, by way of the equivalence of M-Theory
on $S^{1}$ to the Type IIA string theory and the vanishing of
$H_{1}(K3)$, we see that the behaviors are exactly the same.
Hence, we have found that the behaviors of the Type IIA string
theory on $K3$ and M-Theory on $K3$ ``dove-tail" very nicely.
Both obtain points of enhanced gauge symmetry in their respective
moduli spaces from the wrapping of two-branes about collapsing
two-cycles and, by way of the M-Theory/Type IIA equivalence
\WittenI, we see that these are indeed one and the same behavior.

\newsec{
	Enhanced Gauge Symmetry in M-Theory :
	M-Theory on $T^{5} / {\bf Z}_{2} \times S^{1}$
	}

In this section we will examine the enhanced gauge symmetries which
appear in the moduli space of M-Theory on $T^{5} / {\bf Z}_{2} \times
S^{1}$. To do so, however, we will make use of the equivalence
between M-Theory on $T^{5} / {\bf Z}_{2} \times S^{1}$ and the
Heterotic string theory on $T^{5}$. So, we will, in the next
subsection, derive this result.

\subsec{
	M-Theory on $T^{5} / {\bf Z}_{2} \times S^{1}$
	$\sim$
	Heterotic String Theory on $T^{5}$.
}

In this subsection we will prove that M-Theory on $T^{5} / {\bf Z}_{2}
\times S^{1}$ is equivalent to the Heterotic string theory on $T^{5}$.
M-Theory on $S^{1}$, as was proven by Witten \WittenI, is
equivalent to the Type IIA string theory. Hence, M-Theory on
$T^{5} / {\bf Z}_{2} \times S^{1}$ is equivalent to the Type IIA
string theory on $T^{5} / {\bf Z}_{2}$. By, definition \WittenIII,
the Type I$'$ string theory on $T^{5}$ is equivalent to the Type IIA
string theory on $T^{5} / {\bf Z}_{2}$. Furthermore, the Type I$'$
string theory on $T^{5}$ is equivalent, by way of T-Duality, to the
Type I string theory on $T^{5}$. Also, the Type I theory on $T^{5}$
is equivalent by way of Type I/Heterotic ``duality" \WittenIII\ to
the Heterotic string theory on $T^{5}$ \GinspargI.

So, if we trace through all these relations, we find that M-Theory on
$T^{5}/{\bf Z}_{2} \times S^{1}$ is equivalent to the Heterotic
string theory on $T^{5}$. Another way of looking at this is by way
of the equivalence of M-Theory on $T^{5} / {\bf Z}_{2} \times S^{1}$
with the Type IIB string theory on $K3 \times S^{1}$. This
equivalence was established by Dasgupta, Mukhi, and Witten \DasguptaI
\WittenVII. By way of T-Duality \SeibergI\ the Type IIB string theory
on $K3 \times S^{1}$ is equivalent to the Type IIA string theory on
$K3 \times S^{1}$. Furthermore, by way of string-string duality the
Type IIA string theory on $K3 \times S^{1}$ is equivalent to the
Heterotic string theory on $T^{5}$. So, chaining together these
results one again finds that M-Theory on $T^{5} / {\bf Z}_{2} \times
S^{1}$ is equivalent to the Heterotic string theory on $T^{5}$.
Now, let us employ this to examine the points of enhanced gauge
symmetry in the moduli space of M-Theory on $T^{5} / {\bf Z}_{2}
\times S^{1}$.

\subsec{
	M-Theory on $T^{5} / {\bf Z}_{2} \times S^{1}$ :
	Enhanced Gauge Symmetries
}

Now, as we have established that M-Theory on $T^{5} / {\bf Z}_{2}
\times S^{1}$ is equivalent to the Heterotic string theory on $T^{5}$,
let us use this information to examine the points of enhanced gauge
symmetry in the moduli space of M-Theory on $T^{5} / {\bf Z}_{2}
\times S^{1}$. 

As we mentioned in the previous section, the Heterotic string
theory on $T^{5}$, in the previous section we dealt with $T^{3}$
but the remarks apply also to $T^{5}$, possess points of enhanced
gauge symmetry in its moduli space. These points occur in the
moduli space when a given radii of $T^{5}$ becomes self-dual
or one of the sixteen internal radii becomes self-dual.
Hence, by way of the equivalence of M-Theory on $T^{5} / {\bf Z}_{2}
\times S^{1}$ with the Heterotic string theory on $T^{5}$, M-Theory
on $T^{5} / {\bf Z}_{2} \times S^{1}$ should also possess points of
enhanced gauge symmetry in its moduli space. To examine these
points let us ``step-wise" compactify M-Theory on $T^{5} / {\bf Z}_{2}
\times S^{1}$ by compactifying first on $T^{5} / {\bf Z}_{2}$, then
compactifying the resultant theory on $S^{1}$.

As the low-energy limit of M-Theory in eleven-dimensions is
eleven-dimensional supergravity \WittenIII, M-Theory possess a
three-form $A_{3}$ in its spectrum. This three-form \DuffII\ gives
rise to the two-brane and, through duality, the five-brane of
M-Theory in eleven-dimensions. Now, when we compactify
M-Theory on $T^{5} / {\bf Z}_{2}$ the $A_{3} \wedge A_{3} \wedge
F_{4}$ interaction \CremmerI\ of eleven-dimensional supergravity
implies that $A_{3}$ is odd under the action of ${\bf Z}_{2}$. Hence,
$A_{3}$ does not yield any vectors or three-forms when M-Theory
is compactified upon $T^{5} / {\bf Z}_{2}$. The compactification
of $A_{3}$ yields five two-forms and ten-scalars\foot{ One may
easily derive this by looking at the cohomology ring of $T^{5}$.}.
These two-forms \DuffII\ give rise to one-branes in six-dimensions.
Now, upon compactifying the resultant six-dimensional theory on a
further $S^{1}$, these one-branes can wrap about the $S^{1}$. Hence,
as $1 - 1 = 0$ these one-branes yield zero-branes in five-dimensions.
These zero-branes can yield some of the points of enhanced gauge
symmetry for which we are looking.

As we mentioned in the previous section, the mass of a zero-brane
which arises from a $p$-brane wrapping about a $p$-cycle is the
tension of the $p$-brane, $T_{p}$ say, times the area of the
$p$-cycle, $A_{p}$ say. So, if we notate the tension of a one-brane
which arises from compactifying M-Theory on $T^{5} / {\bf Z}_{2}$
as $T_{1}$, then the mass of the five-dimensional zero-brane resultant
from wrapping this one-brane about the $S^{1}$ of radius $R_{6,M}$
is of order $T_{1} R_{6,M}$. Thus, we can easily obtain massless
zero-branes in five-dimensions in the limit $R_{6,H} \rightarrow 0$.
These are some of the gauge particles for which we are looking.

However, one may wonder if there are any other means through which
enhanced gauge symmetries may arise in the compactification of
M-Theory on $T^{5} / {\bf Z}_{2} \times S^{1}$. A quick look at the
Heterotic side will tell us that indeed this is the case. As the Heterotic
theory on $T^{5}$ is equivalent to M-Theory on $T^{5} / {\bf Z}_{2}
\times S^{1}$, we can check to see if we have explored all of the
enhanced gauge symmetries by counting the various ones present
in the Heterotic compactification on $T^{5}$. In the Heterotic
theory \WittenVI\ we have sixteen internal radii to adjust as well
as five others corresponding to the $T^{5}$ upon which the
Heterotic string is compactified. As we found five-two forms and
thus five one-branes in the compactification of M-Theory on
$T^{5} / {\bf Z}_{2}$, these can map to the five different enhanced
gauge symmetries present in the Heterotic theory which arise
as a result of varying the $T^{5}$ radii. However, as the Heterotic
string theory also possess sixteen internal radii we seem at
a bit of a loss. However, we can rectify this by examining
the ``twisted" sector of M-Theory on $T^{5} / {\bf Z}_{2}$.

To do so we will have to take a step back from the task at hand
and examine for a moment the origin of twisted sector states in
the compactification of M-Theory on $T^{5} / {\bf Z}_{2}$. This
particular subject was considered in detail by Dasgupta, Mukhi,
and Witten \DasguptaI \WittenVII. We will simply touch upon
the relevant portions of their work; if the reader is interested in
a more detailed account, they should consult the primary sources
referenced above.

In the work of Dasgupta, Mukhi, and Witten \DasguptaI
\WittenVII\ it was proven that M-Theory on $T^{5} / {\bf Z}_{2}$
is equivalent to the Type IIB string theory on $K3$. So, in
examining the twisted sector of M-Theory on $T^{5} / {\bf Z}_{2}$
we are free to examine either the Type IIB string theory on $K3$
or M-Theory on $T^{5} / {\bf Z}_{2}$. We will, for the moment,
study the Type IIB string theory on $K3$. Type IIB
string theory on $K3$ yields a chiral $N=4$ theory
\WittenVII. The supergravity multiplet of this chiral $N=4$
theory contains a graviton, five self-dual two-forms, and
various gravitinos. This gravitational multiplet is, however,
anomalous in six-dimensions and hence requires the addition
of various other multiplets to cancel the anomaly. The choice
of such multiplets, thankfully, is rather easy. There is actually
only one possible type of matter multiplet in six-dimensional $N = 4$
chiral supergravity, a tensor multiplet. This tensor multiplet
contains an anti-self-dual two-form, five scalars, and various
chiral spinors. Cancellation of the anomaly due to the gravitational
multiplet requires that the six-dimensional theory possess
twenty one such tensor multiplets.

Now, looking at our previous results, M-Theory on $T^{5} /
{\bf Z}_{2}$ possess five two-forms. The two-forms may be split
into self-dual and anti-self-dual portions. Hence, we obtain
five anti-self-dual two-forms and five self-dual two-forms.
Including the other bosonic and fermionic modes in the
compactification of M-Theory on $T^{5} / {\bf Z}_{2}$ one finds
that the five self-dual two-forms are part of the gravitational
multiplet and the five anti-self-dual two-forms are part of
five tensor multiplets. Hence, from the untwisted sector
one obtains just the gravitational multiplet and five tensor
multiplets. However, this spectrum, as it does not contain
the requisite twenty one tensor multiplets, is anomalous.
This is the origin of the ``twisted" states. One must add
to the theory extra states not present in the standard
compactification of M-Theory on $T^{5} / {\bf Z}_{2}$
to yield a non-anomalous theory. The question now is,
exactly how does one add such states to the theory?

This was answered by Witten \WittenVII.  Witten found
that there are an infinite variety of ways to add these states
to the theory, but each of these many ways must satisfy
a set of constraints which we will presently review.

Consider the three-form $A_{3}$ in eleven-dimensional
M-Theory. As this three-form \CremmerI\ is a gauge
field, objects may possess an electric or magnetic charge
with respect to $A_{3}$. Witten found that one could
take the fixed points of $T^{5} / {\bf Z}_{2}$ as magnetic
sources of $A_{3}$ charge. This was, however, subject to
the constraints that the sum of all magnetic charges on $T^{5} /
{\bf Z}_{2}$ must be zero and no fixed point may have
charge less than $-1/2$. Furthermore, the manifold $T^{5} /
{\bf Z}_{2}$ must also possess the $21 - 5 = 16$ tensor
multiplets which are needed for anomaly cancellation.

Now, let us consider an example of such a configuration
which satisfies all of the above constraints. One could
consider placing a magnetic charge $-1/2$ at all $2^{5} = 32$
of $T^{5} / {\bf Z}_{2}$'s fixed points and also placing at
$16$ of the fixed points $16$ different five-branes. A
five-brane is actually a magnetic source of $A_{3}$ charge
\DuffII \WittenVII\ with magnetic charge $1$. The five-brane also
supports a chiral $N=4$ tensor multiplet on its world-volume.
Hence, with this configuration we have satisfied all of the above
requirements. The sum of the magnetic charges is
$32(-1/2) + 16(1) = 0$, no fixed point has charge less than
$-1/2$, and we have the sixteen extra tensor multiplets supported
by the sixteen five-branes. In another more general configuration
one could consider placing a magnetic charge $-1/2$ at each of
the $32$ fixed points and locating the sixteen five-branes at
generic points on $T^{5} / {\bf Z}_{2}$. This again satisfies
all the above constraints, and hence, it is a configuration
which the theory may take.

Now, let us consider what this implies about M-Theory
on $T^{5} / {\bf Z}_{2} \times S^{1}$. As we mentioned earlier,
we have five enhanced gauge symmetries arising from the
five one-branes of M-Theory on $T^{5} / {\bf Z}_{2}$ wrapping
about the collapsing one-cycle $S^{1}$. We need sixteen more
ways to obtain an enhanced gauge symmetry so as to match up
with the Heterotic theory on $T^{5}$. The fact that we also
require sixteen five-branes on $T^{5} / {\bf Z}_{2}$ is, as we
will see, more than just a coincidence.

As we stated previously, each one of the five-branes supports a chiral
$N = 4$ tensor multiplet. This tensor multiplet consists of a
anti-self-dual two-form along with five scalars and various chiral
fermions. The physical interpretation of these scalars can be
ascertained by considering the standard bosonic string.
The bosonic world-sheet theory generically has twenty six scalars,
however, one can rid the theory of two, the dimension of the
world-sheet,  of these twenty-six scalars as a result of the action
of the world-sheet diffeomorphisim group. In a similar light,
consider the world-volume of the five-brane before gauge
fixing. As M-Theory is an eleven-dimensional theory the world-volume
theory would generically have eleven scalars corresponding to the
eleven coordinates. However, one could rid the theory of six of
these scalars by employing the diffeomorphisim group of the
world-volume theory. This would leave the world-volume theory with
five scalars, the exact number we found earlier by other means.
These five scalars, which are part of the tensor multiplet, thus
correspond to physical displacements of the five-brane in the
ambient eleven-dimensional world.

Let us consider a configuration of five-branes and $A_{3}$ charges
in which the the sixteen five-branes reside at sixteen of the thirty
two fixed points of $T^{5} / {\bf Z}_{2}$ and in addition there is
a magnetic charge of $-1/2$ at each of the thirty two fixed points.
This configuration, as we found earlier, satisfies all the relevant
constraints. Now, as we have a physical interpretation for the five
scalars on the five-brane world-volume, we can interpret them in
this configuration. For a given five-brane the values of its five
scalars give its position on the five manifold $T^{5} / {\bf Z}_{2}$.
In this configuration the scalars of the five-branes all take on
values which land the five-branes at the fixed points of $T^{5} /
{\bf Z}_{2}$. However, one could also consider a case in which the
values of the five scalars for any given five-brane vary a bit taking
the five-brane away from its fixed point. As this new situation
also satisfies all the relevant constraints, it is also physical.
So, letting the world volume scalars vary, the generic configuration
consists of a set of sixteen five-branes at generic locations on
$T^{5} / {\bf Z}_{2}$ determined by the values of the various
world-volume scalars. Now, let us consider compactifying this generic
configuration on $S^{1}$.

A five-brane is a six-dimensional object; $T^{5} / {\bf Z}_{2}$ is a
five-dimensional object. In the generic configuration of five-branes
which we are considering, any given five-brane and $T^{5} /
{\bf Z}_{2}$ are transverse. Hence, when we compactify on $S^{1}$
all sixteen of the five-branes must wrap around the $S^{1}$ just
as a result of dimension counting. Now, as the five-brane world-volume
possess a chiral $N=4$ tensor multiplet which supports, among
other fields, an anti-self-dual tensor, the five-brane
world-volume also supports a one-brane \DuffII. This one-brane,
upon compactifying on $S^{1}$,  has states which wrap about this
$S^{1}$ and states which do not wrap about this $S^{1}$.
The states which wrap about this $S^{1}$ appear in the resultant
five-dimensional world as zero-branes and hence, if rendered
massless, the gauge fields for which we are looking. So, again, we
are now in a familiar situation. A $p$-brane wrapping about a
$p$-cycle. As we found earlier, the mass of such a state is the 
tension of the $p$-brane, $T_{p}$ say, times the area of the
$p$-cycle, $A_{p}$ say. In the case at hand, it is the tension of
the one-brane on the five-brane world-volume times the length of
the $S^{1}$ upon which we are compactifying. Hence, as the $S^{1}$
collapses or the one-brane tension vanishes we obtain an enhanced
gauge symmetry in the same manner we found in earlier examples.
However, now our counting works correctly.

As we previously mentioned, M-Theory on $T^{5} / {\bf Z}_{2}
\times S^{1}$ is equivalent to the Heterotic string theory on
$T^{5}$. Varying the radii of the $T^{5}$ as well as the sixteen
internal radii of the Heterotic string theory one can see that
there are twenty one different ways to obtain an enhanced
gauge symmetry in the Heterotic string on $T^{5}$.
Previously, we were unable to find all these different enhanced
gauge symmetries in the M-Theory compactification on
$T^{5} / {\bf Z}_{2} \times S^{1}$; now, with our above comments
on the various five-branes in the theory, we are able to do so.
As we found above, the untwisted sector of M-Theory on
$T^{5} / {\bf Z}_{2}$ possess five two-forms. These five two-forms
give rise to five one-branes. Upon compactifying further on an
$S^{1}$ these five one-branes can wrap about this $S^{1}$ and
give rise to five different ways of obtaining an enhanced gauge
symmetry as the $S^{1}$ collapses. These five different ways
of obtaining an enhanced gauge symmetry correspond to the
five different radii of the $T^ {5}$ factor in the Heterotic string
theory compactification becoming self-dual. Now, on the
Heterotic side we have $21 - 5 = 16$ other ways in which to
obtain an enhanced gauge symmetry. Originally, on the M-Theory
side we did not know the origin of these sixteen extra
enhanced gauge symmetries, now we do. Generically, the
sixteen five branes in the M-Theory compactification on
$T^{5} / {\bf Z}_{2}$ are transverse to $T^{5} / {\bf Z}_{2}$. So,
upon further compactification on $S^{1}$ they wrap about
$S^{1}$. Each of these five-branes also supports a tensor
multiplet which contains a two-form. Thus, each five-brane
has a one-brane living on its world-volume. Upon wrapping
these sixteen five-branes on $S^{1}$, each of these sixteen
one-branes may also wrap about this $S^{1}$. Counting
dimensions, these sixteen wrapped one-branes give rise to
sixteen zero-forms in five-dimensions, and as the $S^{1}$
collapses or the one-brane tension vanishes, these sixteen
one-branes give rise to sixteen ``enhanced" gauge fields. This
mechanism thus provides the sixteen missing ways in which to obtain
an enhanced gauge symmetry in the M-Theory picture. Hence, we now
have an understanding of all twenty one ways in which to obtain
an enhanced gauge symmetry in the M-Theory compactification
on $T^{5} / {\bf Z}_{2} \times S^{1}$.

\subsec{
	Surprises?
	}

In this subsection we will briefly comment on the enhanced
gauge symmetries present in the moduli space of the Type
IIA string theory on $T^{5} / {\bf Z}_{2}$. As M-Theory on
$S^{1}$ is equivalent to the Type IIA string theory \WittenI,
M-Theory on $T^{5}  / {\bf Z}_{2} \times S^{1}$ is equivalent
to the Type IIA string theory on $T^{5} / {\bf Z}_{2}$.  Hence,
the enhanced gauge symmetries in the moduli space of
M-Theory on $T^{5} / {\bf Z}_{2} \times S^{1}$ demand the
existence of enhanced gauge symmetries in the moduli space
of the Type IIA string theory on $T^{5} / {\bf Z}_{2}$.

Now, these enhanced gauge symmetries in the moduli space
of the Type IIA string theory on $T^{5} / {\bf Z}_{2}$ actually
come as no surprise. Type IIA string theory on $T^{5} / {\bf Z}_{2}$
is by definition equivalent to Type I$'$ string theory on $T^{5}$.
Type I$'$ string theory on $T^{5}$ is T-Dual to Type I string
theory on $T^{5}$ \WittenIII. Also, Type I string theory
on $T^{5}$ is equivalent to the Heterotic string theory on
$T^{5}$ by way of the Heterotic/Type I duality \WittenIII.
Hence, as the Heterotic string theory on $T^{5}$ exhibits
points of enhanced gauge symmetry in its moduli space,
so the Type IIA string theory on $T^{5} / {\bf Z}_{2}$ should
also exhibit points of enhanced gauge symmetry in
its moduli space. Notice that we did not rely upon M-Theory
to reach this conclusion. So, our derivation of the existence
of these points of enhanced gauge symmetry from a M-Theory
point-of-view acts as a very strong check on the various relations
between M-Theory and String Theory and the various enhanced gauge
symmetries of both.

\newsec{
	Conclusion
	}

In this article we have examined various points of enhanced gauge
symmetry in the moduli space of M-Theory on $S^{1} \times S^{1} /
{\bf Z}_{2}$, M-Theory on $K3$, and M-Theory on $T^{5} /
{\bf Z}_{2} \times S^{1}$. In each case we found an interesting
interplay of M-Theory and String Theory symmetries. But, our
inquiry left open a few interesting questions which
deserve further study. It would be nice to understand the points
of enhanced gauge symmetry in the strong coupling region of
Figure 1 from a perturbative point-of-view. Also, it would be
nice to explore in a bit more detail the interplay of the one-branes
on the world-volume of the five-brane and their interaction with
spacetime fields. A bit of this interplay between the one-branes
on the five-brane world-sheet and the spacetime fields is given
by our above comments. Also, some of this interplay is
suggested by the comments made by Witten in \WittenVII.
However, it would be interesting to fully understand these
one-branes and how they affect spacetime physics.

\listrefs
\bye